\documentclass[10pt,showpacs,reprint,twocolumn]{revtex4-1}
\usepackage{graphicx}
\usepackage[caption = false]{subfig}
\usepackage{color}
\usepackage{epsfig}
\usepackage{makeidx}
\usepackage{ifpdf}
\usepackage{url}%
\usepackage{cleveref}
\usepackage{float}
\usepackage{bm}
\usepackage[colorlinks=true,linkcolor=blue]{hyperref}%
\expandafter\ifx\csname package@font\endcsname\relax\else
 \expandafter\expandafter
 \expandafter\usepackage
 \usepackage{varioref}
 \expandafter\expandafter
 \expandafter{\csname package@font\endcsname}%
\fi
\begin{document}
\title{Experimental observation of strong coupling effects on the dispersion of dust acoustic waves in a plasma}
\author{P. Bandyopadhyay\footnote[1]{corresponding author.\\ \textit{E-mail address:} pintu@ipr.res.in (P. Bandyopadhyay)},  G. Prasad, A. Sen and P. K. Kaw}
\affiliation{Institute for Plasma Research, Bhat, Gandhinagar - 382428, India }
\begin{abstract}
The dispersion properties of low frequency dust acoustic waves in the strong coupling regime are investigated experimentally in an argon plasma embedded with a mixture of kaolin and $MnO_2$ dust particles. The neutral pressure is varied over a wide range to change the collisional properties of the dusty plasma. In the low collisional regime the turnover of the dispersion curve at higher wave numbers and the resultant region of $\partial\omega/\partial k < 0$ are identified as signatures of dust-dust correlations. In the high collisional regime dust neutral collisions produce a similar effect and prevent an unambiguous identification of strong coupling effects.\\  
\end{abstract}
\maketitle
Dust acoustic waves (DAWs), which are the analogs of ion acoustic waves in a dusty plasma, have received a great deal of attention in recent years. Their linear and nonlinear propagation characteristics have been the subject of a large number of theoretical \cite{rao,verheest,ganguli,hollen,rosen1} and  experimental \cite{barkan,thompson,pieper,fortov,pramanik,prabhakara,merlino} investigations. One of the important and fundamental questions related to their dispersion properties concerns the effect of dust-dust correlations. These correlation effects are deemed to be important because dusty plasmas are usually in the strongly coupled regime on account of the large amount of charge on each dust particle and its relatively low thermal velocity. $\Gamma = \frac{Z_d^2e^2}{4\pi\epsilon_0T_d d}exp[-d/\lambda_p]$,  the ratio of the dust Coulomb energy to the dust thermal energy is widely used as a measure of the amount of coupling. Here, $Z_d$, $d$, $T_d$, $\lambda_p$ denote the dust charge, the interparticle distance, the dust particle temperature and the plasma Debye length respectively. 
The value of this parameter, which is based on the Yukawa model, provides a broad indication of the phase of a dusty plasma i.e. whether it is in a gaseous ($\Gamma<1$), fluid ($1<\Gamma<\Gamma_c$) or solid state ($\Gamma>\Gamma_c$). Here $\Gamma_c$ is the critical value of $\Gamma$ marking a phase transition point. It should be mentioned that a number of recent works \cite{vaulina1,vaulina2} have suggested important modifications to the expression for $\Gamma$ and $\Gamma_c$ near a critical curve. In general, when $\Gamma >>1$ in a dusty plasma, dust-dust correlations can lead to the development of short range order in the system which keeps decaying and reforming in time. Such correlation effects can bring about significant modifications in the collective properties of the strongly coupled dusty plasma \cite{rosen2,kaw,murillo1,murillo2}. The influence of dust-dust correlation on DAWs was addressed in an early experiment by Pieper and Goree \cite{pieper} who studied the propagation of DAWs in a dusty plasma that was close to a crystalline state. The expectation was that the linear dispersion relation of the excited compressive waves would be close to dust-lattice waves - DLWs (evidence of a strongly correlated medium) rather than that of DAWs which are typically found in a gaseous weakly coupled plasma. To their surprise they found that the dispersion properties were well explained by collisionally damped DAWs and showed no resemblance to DLWs. Their experimental findings also lent credence to the applicability of fluid models for describing wave motion in dusty plasmas. Subsequent theoretical \cite{kaw,murillo1,murillo2,kalman1,kalman2} and particle simulation \cite{ohta} studies have highlighted the fact that correlation effects can be important even in the modestly coupled regime in the near liquid state ($\Gamma \sim 10$) and can lead to a turnover effect of the dispersion curve for DAWs at higher wave numbers and create a regime where $\partial\omega/\partial k < 0$. However as has been pointed out in \cite{kaw}, similar effects can also arise due to strong dust-neutral collisions and pose a challenge for an unambiguous experimental detection of dust-dust correlation effects for longitudinal waves. Thus for a clear identification of correlation effects on the dust acoustic wave dispersion, it is
necessary to work in a low collisional regime of the dusty plasma and look for the turnover regime in the dispersion curve. In this paper we report on such an investigation and present experimental evidence of correlation effects on the propagation of dust acoustic waves.\par
\begin{figure*}[ht]
\centerline{\hbox{\psfig{file=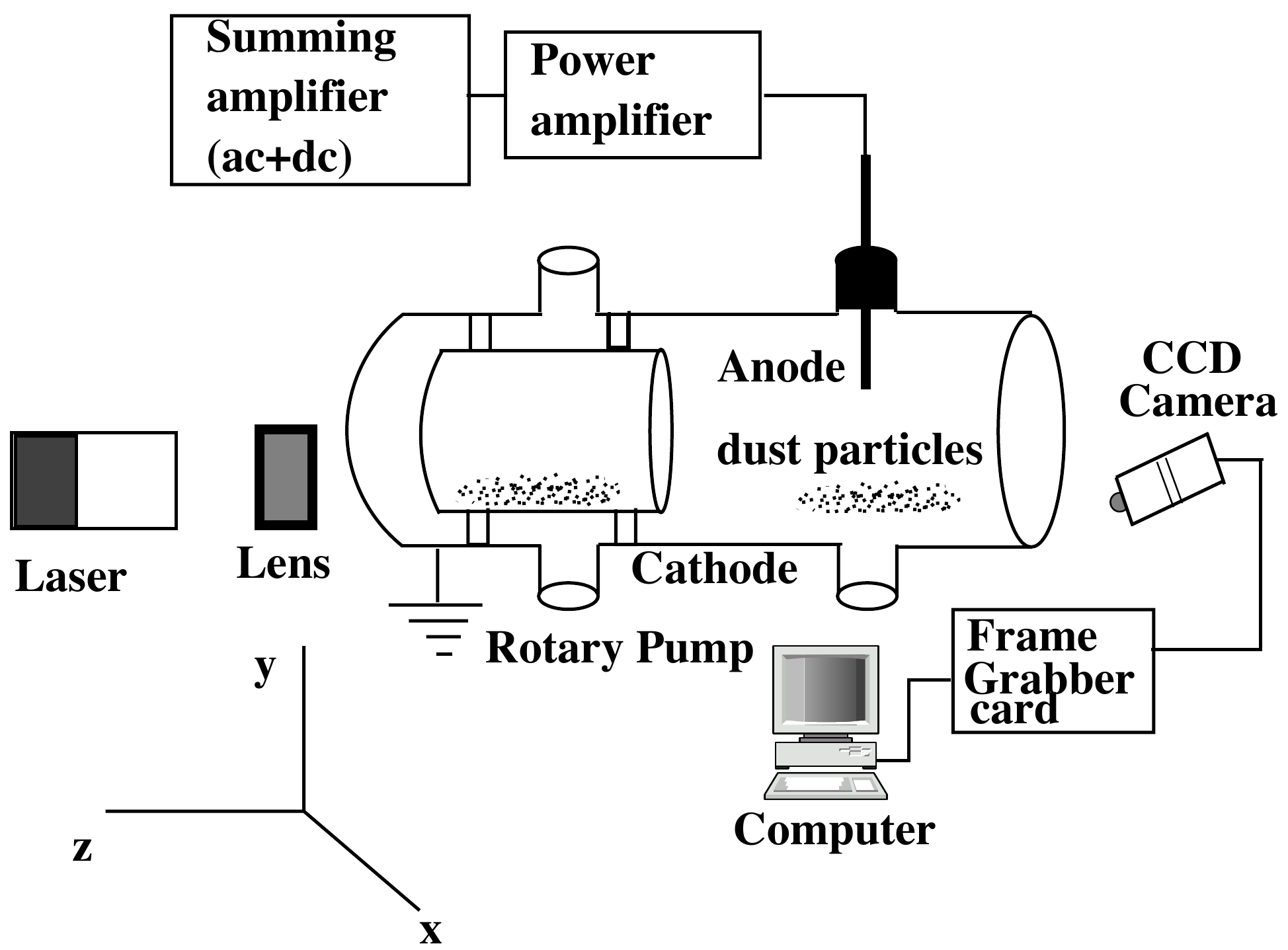,width=.50\textwidth,angle=0}}}
\caption{Schematic of experimental setup.}
\label{qua_2}
\end{figure*}
The schematic of the experimental setup is shown in Fig. 1. The vacuum vessel consisting of a stainless steel (SS) cylinder of 95 cm length and 20 cm diameter, was pumped down to a base pressure of $10^{-3}$ mbar using a rotary pump. Argon gas was then injected by a precision needle valve and discharge was struck between the vacuum vessel (cathode) and a rod shaped anode by applying a DC voltage of $V_{A} = 600$ volts at $P = 1$ mbar pressure. The inner side of the cathode was covered with a thin stainless steel foil to avoid micro-arcs. A mixture of kaolin and  $MnO_2$ was used as dust particles.  The mean particle size was $\sim 31 \mu m$ with a size dispersion of $\sim \pm 3 \mu m$. The size distribution was experimentally determined by SEM measurements of a large sample of the dust particles. To prevent arcing around the dust particles, the latter were kept in a floating SS cylinder housed inside the main chamber, as shown in Fig. 1. Noticeable dust particle accumulation was observed near the cathode sheath region when the neutral gas pressure was reduced gradually by adjusting the leak rate and pumping rate to 0.086 mbar. The applied voltage was then reduced to 400 volt to get a highly dense dust cloud. For such discharge conditions the discharge current was 45 mA. The vertical component of the cathode sheath electric field provides necessary electrostatic force to the particles to levitate against the gravitational force. Likewise the radial horizontal sheath electric field was responsible for the radial confinement of the dust particles against their mutual Coulomb repulsive forces.\par
\begin{figure}[ht]
\centerline{\hbox{\psfig{file=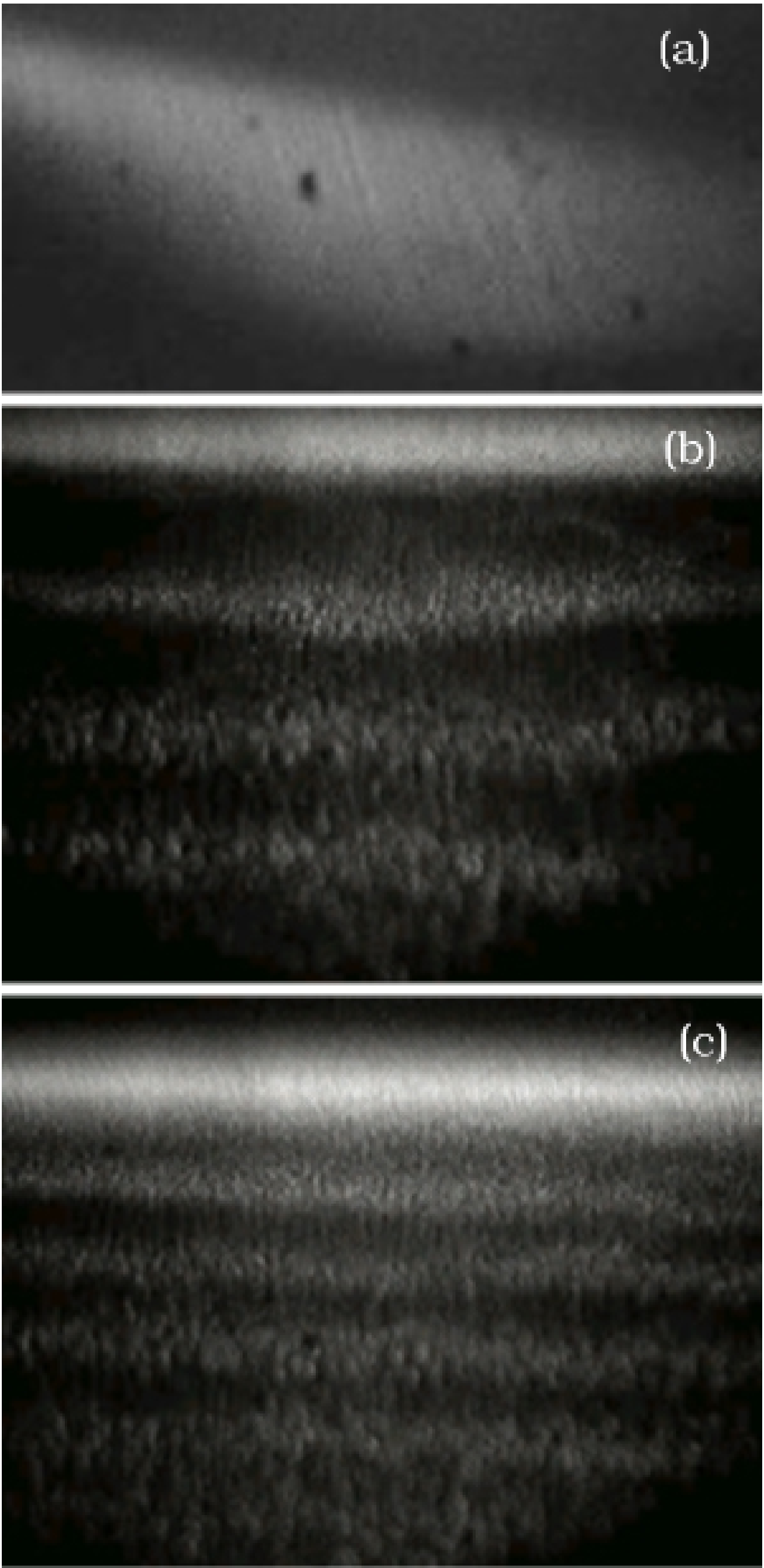,width=.40\textwidth,angle=0}}}
\caption{a) Stable dust cloud b) $\&$ c) DAWs of different wavelengths at f=0.206 Hz and 0.533 Hz respectively.}
\label{qua_2}
\end{figure}
An AC signal (peak to peak voltage $V_{PP}=60$ volt using a signal generator and a power amplifier) was superimposed on the discharge voltage to excite the waves. 
The mean discharge current was 45 mA at 400 volts. The frequency of the applied AC signal was varied from 0 to 2 Hz during the experiment.\par
The levitated dust particles were illuminated by green Nd-Yag diode laser light. The laser light was spread into a sheet of width $\sim 1$ $mm$ in the $x-y$ plane by a cylindrical lens and the forward scattered light from the dust cloud was used to visualize the dust particles. The scattered light from the dust particles was captured using a CCD camera (25 fps) and stored into a computer using a frame grabber card. The ion density ($n_i$) and the electron temperature ($T_e$) of the plasma were measured using a single Langmuir probe prior to the introduction of any dust particles. It should be mentioned that strictly speaking one needs the plasma parameter values inside the sheath region where the dust cloud is confined. However as is well known Langmuir probe measurements inside the sheath region can be quite unreliable \cite{pramanik}. Thus we have measured $T_e$ and $n_i$ in the plasma region outside the sheath and used these values for the sheath region. The electron density ($n_e$) was then estimated from the modified quasineutrality condition that includes the dust density. The average inter-grain distance ($d$) was measured from a number of still video images of the dusty plasma. The dust temperature ($T_d$) was calculated from the mean velocity of the dust particles by tracing single particle trajectories in different frames. The wave length of the driven dust acoustic wave was obtained by a direct measurement of the distance between crests from the still images. To ascertain the propagation direction of the wave propagation an independent set of measurements were made in the vertical plane by changing the orientation of the laser sheet to be in the $y-z$ plane. We did not observe any time varying striations in the plasma cloud for such an orientation indicating that the DAW does not have any significant $k_y$ component and propagates in the $y-z$ plane. Further the wavelength measurements in the horizontal plane were duly corrected to account for the geometrical effects arising from the small but finite angle of inclination ($15^0$) of the CCD camera with respect to the vertical plane. The wavelength measurements were also restricted to the well focused region and a fixed laser-lens combination was retained for all the experimental  observations. \par
\begin{figure}[ht]
\centerline{\hbox{\psfig{file=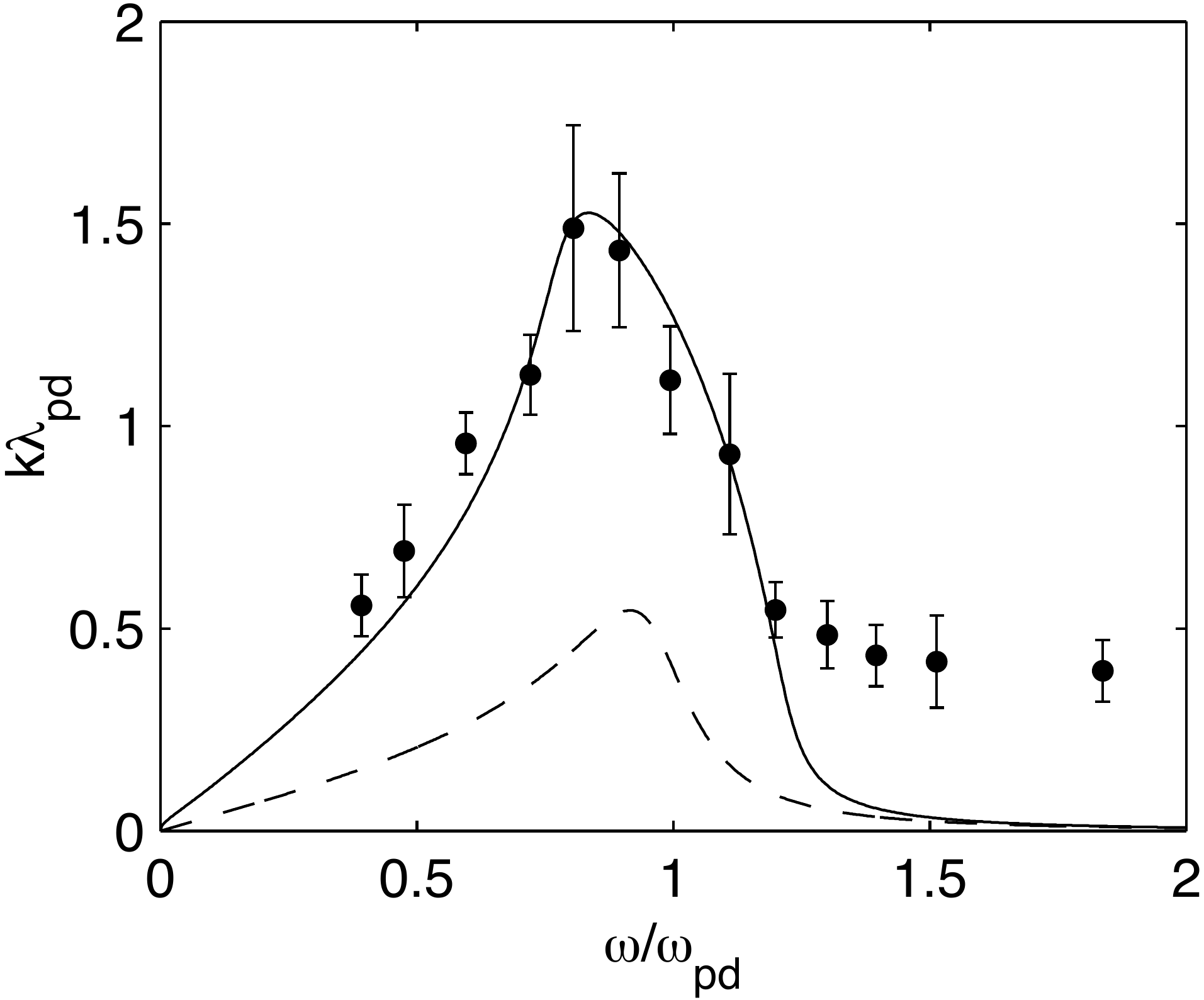,width=.40\textwidth,angle=0}}}
\caption{Experimental (solid circles) and theoretically calculated (solid and dash lines) dispersion curves for DAWs in the low collisional $\nu_{dn}=0.07$ regime. Solid line displays the dispersion relation at experimental parameters and dash line for $\Gamma \sim 0$.}
\label{qua_2}
\end{figure}

\begin{figure}[ht]
\centerline{\hbox{\psfig{file=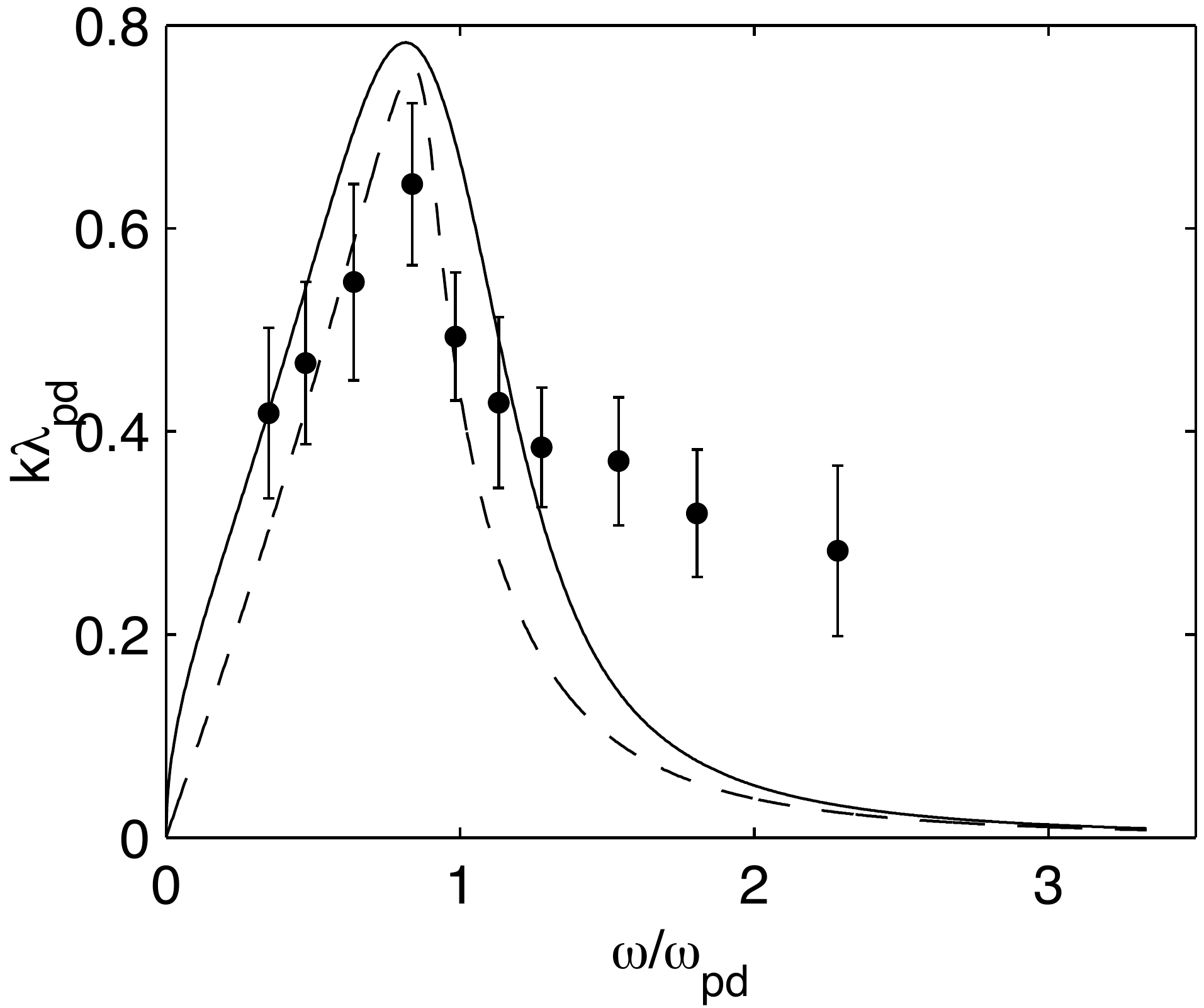,width=.40\textwidth,angle=0}}}
\caption{Experimental (solid circles) and theoretically calculated (solid and dash lines) dispersion curves for DAWs in the high collisional regime $\nu_{dn}=0.6$. Solid line displays the dispersion relation at experimental parameters and dash line for $\Gamma \sim 0$.}
\label{qua_2}
\end{figure}
Initially (at $V_A=400$ Volt, $I_d=45$ mA, $P=0.086$ mbar) a stable dust cloud was formed in the sheath region near the cathode (see Fig. 2(a)). With the imposition of an AC voltage ($V_{pp}=60$ Volt) over the DC discharge voltage, the dust cloud began to display striations which traveled in the horizontal (x$-$z plane) direction. The series of crests and troughs representing the dust acoustic waves are clearly observed. Sample video images are shown in Fig. 2(b) and Fig. 2(c) for different excitation frequencies and are akin to earlier observations of DAWs reported in the literature \cite{barkan,fortov,pieper,pramanik1}. Their wavelengths are determined from the still images with the precautions discussed above.\par
The dispersion characteristics of these DAWs were studied for different neutral pressures which corresponds to different collisionality values. Fig. 3 and Fig. 4 display the experimental results for two different pressure values, namely, $P=0.086$ mbar and $P=0.5$ mbar respectively. The distribution of experimental points in both the cases show a turn-over effect at high wave numbers. In order to distinguish between correlation effects and collisional effects we compare these experimental curves with the analytic dispersion relations provided by model theoretical calculations. The dispersion relation given in \cite{kaw} takes into account both correlation (due to viscoelastic effects) and collisional contributions (due to dust neutral drag) is given below 
\begin{eqnarray}
k^2\lambda_p^2  = \frac{\omega(\omega+i\nu_{dn}) - k^2 \alpha {\lambda_p}^2}
{1-\omega(\omega+i\nu_{dn})+k^2\alpha{\lambda_p}^2}
\label{eqn:dis}
\end{eqnarray}
{\color{black} Here, $\nu_{dn} = \sqrt{8/\pi}n_nT_n/(\omega_{pd}V_{T_n}\rho a)$ is the dimensionless dust neutral collision frequency (i.e. dust neutral frequency normalized by the dust plasma frequency $\omega_{pd}$)} , and $n_n, T_n, V_{T_n}$  are the neutral density, neutral temperature and neutral thermal velocity respectively. Further $\alpha = \gamma_d\mu_d \lambda_d^2/{\lambda_p}^2$ where $\gamma_d$ is the adiabatic index and $\mu_d$ is the coefficient of compressibility. The term proportional to $\alpha$ represents the strong coupling contribution arising through the modification in $\mu_d$. The viscoelastic model \cite{kaw} gives
\begin{eqnarray}
\mu_d = \frac{1}{T_d}\times\left(\frac{\partial P}{\partial n} \right)_T = 1 + \frac{u(\Gamma)}{3} + \frac{\Gamma}{9}\frac{\partial u(\Gamma)}{\partial \Gamma}
\end{eqnarray} 
with
\begin{eqnarray}
u(\Gamma) = -0.89\Gamma + 0.95\Gamma^{1/4} + 0.19\Gamma^{-1/4} - 0.81.
\end{eqnarray} 
For the low pressure regime (P=0.086 mbar) the measured dusty plasma parameters are $T_d=5 eV$, $n_e \sim 5\times10^{13}m^{-3}$, $n_i =3\times10^{14} m^{-3}$ and $d=150 \mu$m. 
The ion temperature is taken to be $T_i=0.03$. These measured quantities give a plasma Debye length $\lambda_p = 190$ $\mu$m. The wave number values are normalized to $\lambda_p$ and the frequencies are normalized to $\omega_{pd}$. The latter is used a free parameter and determined from the best curve fit to the experimental points. In the present case $\omega_{pd}=3.3$ rad/s. From the measured quantities and 
$\omega_{pd}$ we get the dust charge $Q_d$ to be $5.8 \times 10^3e$, $\Gamma =29$, $\alpha \sim -0.05$ and $\nu_{dn}=0.07$. In Fig. 3 we have plotted the dispersion relation (\ref{eqn:dis}) (solid curve) for the above values. For comparison we have also plotted the same dispersion curve without correlation effects (dashed curve).  It is clear that for a proper fit to the experimental data points, it is necessary to include correlation contributions in the dispersion relation indicating that they play a dominant role in this regime. The dashed curve infer purely collisional effects taken into account. It suggests that the collisional effect themselves are not adequate to explain the observed magnitudes of the dust acoustic wave frequencies and wave numbers observed in the experiments.\par
As the neutral pressure rises the collisional effects begin to dominate. In the high pressure regime (P = 0.5 mbar) we have  $\nu_{dn}=0.6$. To excite the DAWs in this regime the applied DC voltage was raised to 500 volt and the AC component was also increased to 100 volt. The mean discharge current increased to 60 mA due to the increase of neutral gas pressure and applied voltage. The corresponding theoretical and experimental dispersive characteristics are shown in Fig. 4. The theoretical curve was fitted using $\omega_{pd}=3$ rad/sec. The dusty plasma parameters for present pressure value were $d=90 \mu$m, $T_d=4$eV, $n_i \sim 7\times10^{14}/m^3$, $n_e=7\times 10^{13}/m^3$, $Q_d = 2 \times 10^3e$ and $\Gamma = 6.4$. The decrease of $\Gamma$ in the present case can be attributed  to the decrease in the dust charge. The decrease in dust temperature for a marginal increase in the pressure has also been reported in \cite{pieper}. The decrease in $\Gamma$ leads to a decrease in the contribution from the correlation term and the high neutral pressure makes the collisional contributions to predominate. This is evident from the two theoretical curves in Fig. 4 where the neglect of the $\alpha$ term (dashed curve) does not make a substantial difference and both curves are seen to be close to the experimental points. In this regime, therefore, it is difficult to experimentally differentiate between the collisional and correlation effects. It should be mentioned here that the above dispersion curves can be influenced by other factors such as the size dispersion of the dust particles. We have used the mean dust size for estimating the dust charge as well as other theoretical entities like the dispersion relation, coupling parameter etc. Since the dispersion is not very broad the percentage changes in these quantities are quite small - notably within the experimental error bars of the measured quantities. Hence they do not significantly affect our principal experimental findings related to the effect of correlations on the wave dispersion properties. \par
To conclude we have presented experimental evidence of dust correlation effects on the dispersive characteristics of dust acoustic waves. The experimental dispersion curve shows good agreement with theoretical relations obtained using a generalized hydrodynamics model. Correlation effects can be clearly distinguished from collisional effects in the low pressure regime. In order to further verify experimental findings we have studied the dispersion characteristics in both the high and low pressure regimes. Comparison with theoretical relations in the two regimes demonstrate that the dispersion curves show a clear signature of correlation effects in the low collisional regime. In the high collisional regime dust neutral collisions lead to similar changes in the dispersion curve and make an unambiguous experimental identification of correlation effects quite difficult. 

\end{document}